\documentclass{article}
\usepackage{cite} 
\usepackage{amsmath} 
\usepackage{mathtools} 
\usepackage{bm} 
\usepackage{float}
\usepackage{soul}
\usepackage[dvipsnames]{xcolor}
\usepackage[hidelinks]{hyperref}
\usepackage[font={small}]{caption}
\usepackage{authblk}

\let\citeleft=(
\let\citeright=)

\newtagform{brackets}{[}{]}
\usetagform{brackets}

\begin{document}

\pdfinfo{
   /Author (AUTHORS)
   /Title (TITLE)
}

\makeatletter 
\renewcommand\@biblabel[1]{#1.} 
\makeatother

\definecolor{red}{rgb}{1,0,0}


\def\ebf{{\mathbf{e}}}
\def\rbf{{\mathbf{r}}}
\def\sbf{{\mathbf{s}}}
\def\xbf{{\mathbf{x}}}
\def\ybf{{\mathbf{y}}}
\def\fbf{{\mathbf{f}}}
\def\pbf{{\mathbf{p}}}

\def\rbfhat{{\widehat{\mathbf{r}}}}
\def\sbfhat{{\widehat{\mathbf{s}}}}
\def\pbfhat{{\widehat{\mathbf{p}}}}

\def\Ical{{\mathcal{I}}}
\def\Dcal{{\mathcal{D}}}
\def\Lcal{{\mathcal{L}}}
\def\Mcal{{\mathcal{M}}}

\def\rbftilde{{\tilde{\mathbf{r}}}}
\def\sbftilde{{\tilde{\mathbf{s}}}}
\def\pbftilde{{\tilde{\mathbf{p}}}}

\def\thetabm{{\bm{\theta}}}
\def\nubm{{\bm{\nu}}}

\title{\vspace{-2cm} Deep learning using a biophysical model for Robust and Accelerated Reconstruction (RoAR) of quantitative, artifact-free and denoised $R_2^\ast$ images}

\author[1]{Max~Torop}
\author[2]{Satya~VVN~Kothapalli}
\author[1]{Yu~Sun}
\author[3]{Jiaming~Liu}
\author[2]{Sayan~Kahali}
\author[2]{Dmitriy~A.~Yablonskiy}
\author[1,3]{Ulugbek~S.~Kamilov}

\affil[1]{\small Department of Computer Science and Engineering, Washington University in St.~Louis, St.~Louis, MO 63130, USA}
\affil[2]{\small Department of Radiology, Washington University in St.~Louis, St.~Louis, MO 63110, USA}
\affil[3]{\small Department of Electrical and Systems Engineering, Washington University in St.~Louis, St.~Louis, MO 63130, USA}
\maketitle

\vfill
\noindent
\textit{Running head:} Deep learning for robust and fast $R_2^\ast$ reconstruction

\noindent
\textit{Address correspondence to:} \\
Ulugbek~S.~Kamilov, One Brookings Drive, Campus Box 1045, St.~Louis, MO 63130, USA.\\
Email: \url{kamilov@wustl.com}\\
Dmitriy~A.~Yablonskiy, Mallinckrodt Institute of Radiology,  4525 Scott Ave, Room 3216, St.~Louis, MO 63110, USA.\\
Email: \url{yablonskiyd@wustl.edu}

\noindent
This work was supported in part by NSF award CCF-1813910, NIH/NIA grant  R01AG054513, Marilyn Hilton Award for Innovation in MS Research, Amazon Web Services Cloud Credits for Research program, and NVIDIA Corporation with the donation of the Titan Xp GPU for research.

\noindent
Approximate word count: 250 (Abstract)  3,500 (body)\\

\noindent
Accepted to \emph{Magnetic Resonance in Medicine} as a Rapid Communication.\\

\clearpage

\section*{Abstract}

\noindent
\textbf{Purpose}: To introduce a novel deep learning method for Robust and Accelerated Reconstruction (RoAR) of quantitative and $B0$-inhomogeneity-corrected $R_2^\ast$ maps from multi-gradient recalled echo (mGRE) MRI data. 

\noindent
\textbf{Methods}: RoAR \emph{trains} a convolutional neural network (CNN) to generate quantitative $R_2^*$ maps free from field inhomogeneity artifacts by adopting a self-supervised learning strategy given (a) mGRE magnitude images, (b) the  biophysical model describing mGRE signal decay, and (c) preliminary-evaluated F-function accounting for contribution of macroscopic $B0$ field inhomogeneities. Importantly, no ground-truth $R_2^\ast$ images are required and F-function is only needed during RoAR training but not application.

\noindent
\textbf{Results}: We show that RoAR preserves all features of $R_2^\ast$ maps while offering significant improvements over existing methods in computation speed (seconds vs.\ hours) and reduced sensitivity to noise. Even for data with  SNR=5  RoAR produced  $R_2^\ast$  maps  with  accuracy  of  $22\%$ while  voxel-wise  analysis accuracy was $47\%$. For SNR=10 the RoAR accuracy increased to $17\%$ vs.\ $24\%$ for direct voxel-wise analysis.

\noindent
\textbf{Conclusion}: RoAR is trained to recognize the macroscopic magnetic field inhomogeneities directly from the input magnitude-only mGRE data and eliminate their effect on $R_2^*$ measurements. RoAR training is based on the biophysical model and does not require ground-truth $R_2^\ast$ maps. Since RoAR utilizes signal information not just from individual voxels but also accounts for spatial patterns of the signals in the images, it reduces the sensitivity of $R_2^\ast$ maps to the noise in the data. These features plus high computational speed provide significant benefits for the potential usage of RoAR in clinical settings.

\noindent
\textbf{Keywords}: MRI, gradient recalled echo, $R_2^\ast$ mapping, self-supervised deep learning.

\clearpage

\section*{Introduction}
\label{sec:introduction}

Multi-Gradient-Recalled-Echo (mGRE) sequences are often used for different MRI applications with some even completing initial multi-center testing (e.g.~\cite{ropele2014multicenter, weiskopf2013quantitative}). Since the major goal of mGRE approaches is to produce quantitative metrics related to biological tissue microstructure, it is important to reduce the sensitivity of these metrics to the adverse effects of $B0$ magnetic field inhomogeneities~\cite{yablonskiy1998quantitation, fernandez2000postprocessing, hernando2012r, yablonskiy2013voxel}, and noise in the MRI signal. It is also important to develop metrics-generating algorithms that are fast and robust for direct implementation on MRI scanners.  

Recent methods of image analysis in MRI are increasingly based on deep learning (e.g.~\cite{Han.etal2017, Lee.etal2018, Aggarwal.etal2018, Lee.etal2019}). The current paradigm is based on training the weights of an artificial neural network (ANN) over a dataset in order for the network to produce an accurate estimate of the desired images. Generally, such training is done in a supervised fashion by collecting a large dataset of input signals and corresponding ``ground-truth'' images generated by processing data on a voxel-by-voxel analysis. The key benefit of using deep learning over the traditional voxel-by-voxel  analysis lies in its ability to deal with noisy input and superior computational speed.  This was recently demonstrated on  a related problem of estimating the oxygen extraction fraction (OEF) maps, where an ANN, with a single hidden layer, was trained and applied voxel-wise to produce the desired quantitative parameters given the mGRE signal input~\cite{Domsch.etal2018, Hubertus.etal2019}. The ground-truth datasets for supervised training in~\cite{Domsch.etal2018, Hubertus.etal2019} were generated using simulated signals based on the analytical model~\cite{Yablonskiy.Haacke1994, ulrich2016separation}.

In this paper, we present a novel method, Robust and Accelerated Reconstruction (RoAR)\footnote{The code and the trained models presented in this paper are openly available at \url{https://github.com/wustl-cig/RoAR}.}, that is based on self-supervised training of a deep convolutional neural network (CNN). RoAR does not need ground truth quantitative images (maps) for training; instead, it is trained using only actual mGRE signals, the pre-calculated contribution of magnetic field inhomogeneities to the mGRE signal decay (described in terms of a factor $F(t)$ \cite{yablonskiy1998quantitation} in the mGRE signal model), and our knowledge of the analytical biophysical model connecting the mGRE signal with biological tissue microstructure. Herein we exemplify our analysis by demonstrating the efficiency and robustness of the RoAR method for generating quantitative maps of the mGRE signal transverse relaxation metric $R_2^\ast = 1/T_2^\ast$.
 
The key advantage of RoAR is that it eliminates the need to explicitly characterize the macroscopic magnetic field inhomogeneities to produce $B0$-inhomogeneity-corrected $R_2^\ast$ maps, only requiring $F(t)$ during training. This means that the trained CNN can be directly applied on the mGRE magnitude images without the need to produce images of the $B0$ field or compute $F(t)$. Additionally, by learning statistical dependencies in neighboring voxels, the CNN acts as an imaging prior that can regularize ill-posed imaging problems. Several recent studies have in fact shown that CNNs can serve as excellent priors for a wide range of imaging problems~\cite{VanVeen.etal2018, Sun.etal2018, Ulyanov.etal2018, Liu.etal2019, Sun.etal2019b, Mataev.Elad2019}. 

Our results on experimental and synthetic data  show that  the  proposed  CNN-based $R_2^\ast$ computation method not only reduces the computation time by several orders of magnitude compared with the direct voxel-wise computation, but also significantly improves the computation outcome quality, thanks to the powerful regularization ability of our trained deep CNNs.

\section*{Methods}
\label{sec:methods}

The mGRE signal from a single voxel can be expressed as~\cite{yablonskiy1998quantitation}:
\begin{equation}
\label{Eq:Model}
S(t) = S_0 \cdot \exp(-R_2^\ast \cdot t - i \omega t) \cdot F(t),
\end{equation}
where $t$ denotes the gradient echo time, $S_0 = S(0)$ is the signal intensity at $t = 0$ and $\omega$ is a local frequency of MRI signal. The complex  valued function $F(t)$ in Eq.~\eqref{Eq:Model} describes the effect of macroscopic magnetic field inhomogeneities on the mGRE signal. The failure to account for such inhomogeneities is known to bias and corrupt the recovered $R_2^\ast$ maps. In this paper we use the Voxel Spread Function (VSF) approach~\cite{yablonskiy2013voxel} for calculating $F(t)$. The VSF method takes advantage of both the magnitude and phase of mGRE data from the same mGRE scan that is used for generating $R_2^\ast$ maps. In a standard approach, the $R_2^\ast$ maps, $\omega$ maps, and $S_0$  are jointly estimated from 3D mGRE signals acquired at different echo times $t$ by fitting Eq.~\eqref{Eq:Model} with pre-calculated $F(t)$ on a voxel-by-voxel basis to experimental data by applying the non-linear least squares (NLLS) analysis. In this  paper we propose a computational algorithm that is based on training a convolutional neural network. 

\subsection*{RoAR: Architecture and training}


\begin{figure}[t]
\centering \includegraphics[width=0.9\linewidth]{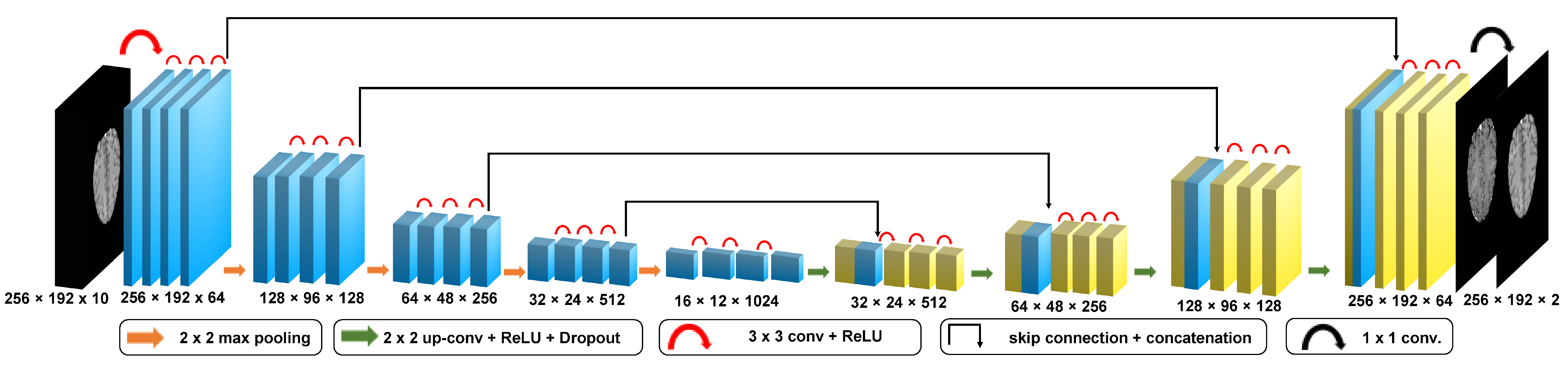} %
\caption{CNN model with a 10-channel input for the mGRE data $\sbf = (\sbf_1, \dots, \sbf_{10})$ and 2-channel output for computed maps of $S_0$ and $R_2^\ast$ ($\pbf = (\sbf_0, \rbf_2^\ast)$). Our model processes data from individual spatial slices extracted from 3D MRI data. The 3D image of the whole brain is obtained by concatenating the outputs of the CNN applied slice-by-slice.}
\label{Fig:CNN}
\end{figure}


Figure~\ref{Fig:CNN} presents the details of our CNN model, which is based on the popular 2D U-Net architecture~\cite{Ronneberger.etal2015}. The U-Net has been extensively used in medical image analysis and relies on a multi-scale decomposition, based on max-pooling, to make the effective size of its filters in the middle layers larger than that of the early and late layers~\cite{Jin.etal2017a}. Such multi-scale structure leads to a large receptive field of the CNN that has been shown to be effective for removing globally spread imaging artifacts typical in MRI~\cite{Han.etal2017}.

Let $\sbf = \Mcal(\pbf ; \fbf)$ denote the magnitude of the signal model in Eq.~\eqref{Eq:Model}, applied to a single 2D spatial slice extracted from the full 3D MRI data. Here $\fbf$ represents the magnitude of pre-computed $F(t)$ values, stored in an array. We represent the magnitudes of the measured mGRE signal of $N$ echo times as the vector
\begin{equation}
\label{Eq:MeasuredData}
\sbf = (\sbf_1, \dots, \sbf_N),
\end{equation}
and represent the corresponding absolute value of $S_0$ and true $R_2^\ast$ maps as another vector
\begin{equation}
\label{Eq:DesiredData}
\pbf = (\sbf_0, \rbf_2^\ast).
\end{equation}
Each vector $\sbf_n$ in $\sbf$ denotes a vectorized 2D image representing the magnitude of the data for one of the echo times.

Let $\pbfhat = \Ical_\thetabm(\sbf)$ denote our model, implemented using our deep CNN architecture, that computes values $\pbfhat$ of the unknown true values of $\pbf$ given the mGRE signal $\sbf$. The vector $\thetabm$ denotes the trainable set of weights in the CNN. In order to increase the expressive power of the network~\cite{Krizhevsky.etal2012}, we rely on multichannel filters, which lead to multiple feature maps at each layer. In our data, the CNN takes $\sbf$ as its 10-channel input and produces $\pbfhat = (\widehat{\sbf}_0, \widehat{\rbf}_2^\ast)$ as its 2-channel output. The volumetric image of the whole brain is obtained by applying the model slice by slice. Note the difference between our deep convolutional architecture from that of recent approaches~\cite{Domsch.etal2018, Hubertus.etal2019} that apply a fully connected ANN voxel-by-voxel to map the mGRE signal to the desired quantitative parameters. These architectures consist of one input layer, one hidden layer with 10 nodes, and one output layer. The convolutional structure of our architecture allows it to process the whole slice of the volumetric data, thus taking into account complex statistical relationships between pixels and echo times.


\begin{figure}[t]
\centering \includegraphics[width=0.9\linewidth]{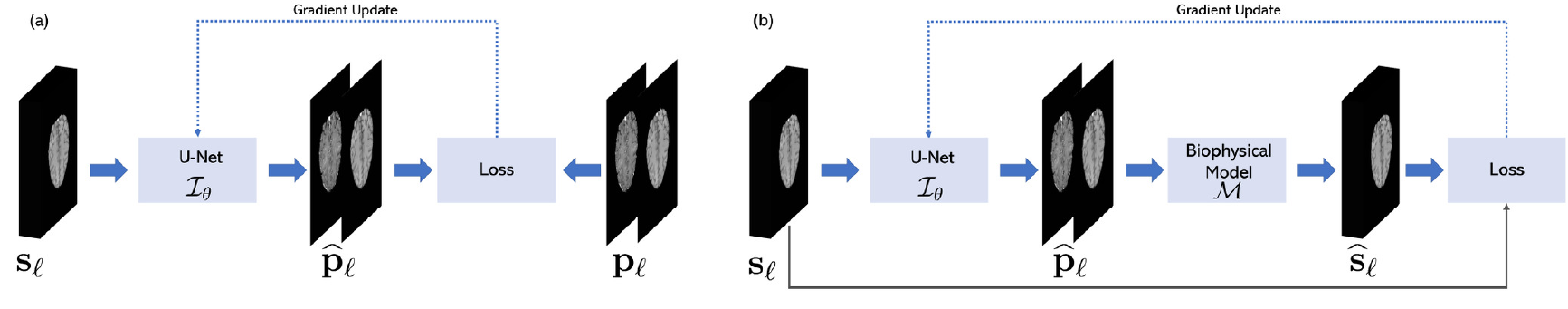} %
\caption{Comparison of two approaches for model training. (a) In the standard supervised approach, the model $\Ical_\thetabm$ is optimized for the loss in the image domain, so that $\pbfhat = \Ical_\thetabm(\sbf)$ is close to the corresponding ground-truth images $\pbf$. (b) In the proposed self-supervised approach, only access to the measurements $\sbf$ and the biophysical model $\Mcal$ is assumed. The loss is formulated in the measurement domain, and the model is trained so that $\sbfhat = \Mcal(\Ical_\thetabm(\sbf); \fbf)$ is close to $\sbf$}
\label{Fig:SelfSupervised}
\end{figure}


The traditional \emph{supervised learning}, illustrated in Figure~\ref{Fig:SelfSupervised}(a), is carried out by minimizing the empirical loss over a training set consisting of $L$ slices $\{(\sbf_\ell, \pbf_\ell)\}_{\ell = 1, \dots, L}$, as follows
\begin{equation}
\label{Eq:Supervised}
\min_{\thetabm} \sum_{\ell = 1}^L \Lcal(\Ical_\thetabm(\sbf_\ell), \pbf_\ell),
\end{equation}
where $\Lcal$ measures the discrepancy between the estimation $\pbfhat_\ell = \Ical_\thetabm(\sbf_\ell)$ generated by the CNN and the ground-truth $\pbf_\ell$. Typical choices for $\Lcal$ include the Euclidean and the $\ell_1$ distances. In practice, the minimization problem is solved by using stochastic gradient-based optimization algorithms such as Adam~\cite{Bottou2012, Kingma.Ba2015}. Once the optimal set of parameters $\thetabm^\ast$ has been learned on the training data, the operator $\Ical_{\thetabm^\ast}$ is applied to solve the computation problem on previously unseen data. The major limitation of the traditional learning paradigm is that it requires large amounts of ground-truth data $\{\pbf_\ell\}$ for training, which can be challenging to obtain in practice. One strategy, adopted by~\cite{Domsch.etal2018, Hubertus.etal2019}, is to synthetically generate such data in individual voxels using the signal model  (e.g. Eq.~\eqref{Eq:Model}). However, this strategy can be sensitive to the mismatch between the \emph{actual} statistical distribution of entries in $\pbf_\ell$ (which might have spatial statistical dependencies) and the \emph{assumed} distribution used in simulation (often independent in space).

The key idea of RoAR is to use \emph{self-supervised learning}, illustrated in Figure~\ref{Fig:SelfSupervised}(b), to train the parameters $\thetabm$ of the  model $\Ical_\thetabm$. 
The idea of using self-supervised learning has recently gained popularity in several distinct imaging applications for addressing the lack of ground-truth training data~\cite{DBLP:journals/corr/abs-1910-09116,chen2019self,kim2019mumford,senouf2019self,lu2019learning}. Recent work in MRSI spectral quantification has seen the integration of CNN's with physical models as a means to avoid dependence on ground-truth labels~\cite{gurbanietal}. The self-supervised learning in RoAR is enabled through our knowledge of the analytical biophysical model connecting the mGRE signal with biological tissue microstructure. Given a set of mGRE measurements $\{\sbf_\ell\}$ and a set of arrays containing corresponding F-function values $\{\fbf_\ell\}$ calculated from the same dataset using the VSF method, our training approach can be formalized as the following optimization problem
\begin{equation} 
\label{Eq:SelfSupervised}
\min_{\thetabm} \sum_{\ell = 1}^L \Lcal\left( \Mcal(\Ical_\thetabm(\sbf_\ell); \fbf_\ell), \sbf_\ell\right).
\end{equation}
Note that the loss function $\Lcal$ in Eq.~\eqref{Eq:SelfSupervised} operates \emph{exclusively} in the measurement space and does \emph{not} require any ground-truth data $\{\pbf_\ell\}$.

Solving the above optimization problem for the $\ell$th data element yields $\pbfhat_\ell = \Ical_\thetabm(\sbf_\ell)$, which acts as a latent image in the intermediate stage for our optimization, as shown in Figure~\ref{Fig:SelfSupervised}(b). Intuitively, our method is searching for images $\pbfhat_\ell$ that are parameterized by $(\thetabm, \{\sbf_\ell\})$ that best explain the measured mGRE signal dataset $\{\sbf_\ell\}$. This strategy is called self-supervised because the measurements themselves provide the supervision to solve the computational problem by exploiting the signal model $\Mcal$ and the prior induced by the CNN. 

\subsection*{Denoising with RoAR}
In addition to RoAR's ability to generate $R_2^\ast$ maps free of field inhomogeneity artifacts based on actual experimental data, RoAR can also be trained to decrease the influence of noise on RoAR-generated $R_2^\ast$ maps. This can be achieved by generating synthetic mGRE data $\{\sbf_\ell\}$ from high SNR data and adding different levels of noise to produce datasets $\{\sbftilde_\ell\}$, so that $\sbftilde_\ell = \sbf_\ell + \ebf_\ell$, where $\ebf_\ell$ denotes the noise and train RoAR by solving the following optimization problem

\begin{equation} 
\label{Eq:NoisySelfSupervised}
\min_{\thetabm} \sum_{\ell = 1}^L \Lcal\left( \Mcal(\Ical_\thetabm(\sbftilde_\ell); \fbf_\ell), \sbf_\ell\right).
\end{equation}
Here, it is important to emphasize that in this approach RoAR requires synthetic data to be generated from high SNR data \emph{only} during training. As corroborated by our results, the trained model yields excellent results at test time, even on previously unseen data with high noise. As described below, the training is done by training the CNN with data containing varying amounts of noise. 

\subsection*{In vivo brain dataset}

For validating our method, we used previously published~\cite{wen2018genetically} brain image data collected from 26 healthy volunteers (age range 26-76) using a Siemens 3T Trio MRI scanner and a 32-channel phased-array head coil. Studies were conducted with the approval of the local IRB of Washington University. All volunteers provided informed consent. The data was obtained using a 3D version of the mGRE sequence with 10 gradient echoes followed by a navigator echo~\cite{Wen.etal2015} used to reduce artifacts induced by physiological fluctuations during the scan. Sequence parameters were flip angle FA $= 30^\circ$, voxel size of $1 \times 1 \times 2$ mm$^3$, first echo time $t_1 = 4$ ms, echo spacing $\Delta t = 4$ ms (monopolar readout gradients), repetition time TR $= 50$ ms, and the total imaging time for each acquisition was around 10 min. 9 of the 26 volunteers were scanned twice, at different times, making for a set of 35 MRI data.

After applying the Fourier transform to the k-space, data from different channels were combined for each voxel to give a single mGRE signal $S(t)$ as described in~\cite{Luo.etal2012}. The effects of macroscopic magnetic field inhomogeneities were taken into account by including the function $F(t)$ in the training of the CNN, Eq~\eqref{Eq:Model}, with $F(t)$ pre-estimated using the voxel spread function (VSF) approach~\cite{yablonskiy2013voxel}.

\subsection*{Performance evaluation}
\label{sec:PerformanceEvaluation}
We evaluated our method on both experimentally measured and synthetic data and compared the results to the traditional voxel-wise NLLS approach. First, we have directly trained RoAR on in vivo data to test  its performance on actual experimental MRI data. The ability of RoAR to correct for noise is validated and quantified by training our model on noisy synthetic data with different noise levels. All neural networks were trained on a GeForce GTX 1080 Ti GPU (NVIDIA Corporation, Santa Clara, CA, USA), and implemented in TensorFlow~\cite{tensorflow2015-whitepaper}, using the Adam optimizer to minimize the Euclidean distance. Prior to being split into slices for training and testing, each mGRE dataset is normalized to ensure that RoAR works with a variety of scanners, parameters, and intensity values. The normalization consists of dividing each value of the signal in a given mGRE dataset by the mean brain voxel signal value in echo 1 of that dataset. This allows RoAR to make $R_2^\ast$ estimations without additional retraining regardless of the initial intensity scale in the mGRE data. RoAR estimates a normalized $S_0$ map from this normalized data and $R_2^\ast$ maps that are not affected by the initial signal intensity. $S_0$ estimations may be rescaled to the initial intensity level using the normalizing value.

Before running NLLS on in vivo MRI data, the VSF method~\cite{yablonskiy2013voxel} was used to find $F(t)$ values at each voxel and each echo time $t$. Here $F(t)$ accounts for the adverse effect of macroscopic magnetic field inhomogeneities on a signal from a given voxel, at echo time $t$ and is essential for true $R_2^\ast$ evaluation. At each iteration of the regression the signal model from Eq.~\eqref{Eq:Model} is parameterized by both the $S_0$ and $R_2^\ast$ estimations as well as given $F(t)$. This accounts for the contribution of field inhomogeneities in the input and results in $R_2^\ast$ estimations free of $B0$ inhomogeneity artifacts. A brain extraction tool, implemented in the Functional Magnetic Resonance Imaging of the Brain Library(FMRIB), was used to mask out both skull and background voxels in all MRI data~\cite{pechaud2006bet2}. NLLS was run over only the set of unmasked voxels, optimizing for 400 iterations at each spatial point. This method was implemented in MatLab R2018b (MathWorks, Natick, MA). The results of this method are then compared to those of a RoAR instance trained on in vivo data for approximately $7$ hours. For comparison, a traditional supervised approach was also implemented, using in vivo data as input and NLLS predictions as ``ground truth'' $R_2^\ast$ to minimize Eq.~\eqref{Eq:Supervised}. This supervised approach was trained for approximately $1$ hour, but required many prior hours for pre-calculating the F-function and producing explicit $R_2^\ast$ maps using the NLLS approach.

To further decrease the influence of noise on RoAR-generated $R_2^\ast$ maps and to test the efficacy of this method, we elected to generate synthetic 3D mGRE signals that are based on real mGRE data to serve as a ``ground truth''. This approach ensures that synthetic data reflects the spatial statistical dependencies seen in real mGRE data. To do this we first run the NLLS method on a high SNR in vivo MRI to get maps of $S_0$  and $R_2^\ast$. These maps, together with the pre-calculated F-function, are used to parameterize the magnitude of the signal model shown in Eq.~\eqref{Eq:Model}, generating signal on a voxel-by-voxel basis. Including the F-function in generated data adds realistic macroscopic magnetic inhomogenieties to the synthetic data. These generated signals constitute a full synthetic 3D mGRE dataset. The 3D $S_0$ and $R_2^\ast$  maps  used  to  generate  a  given synthetic  MRI  can  be  thought  of  as  its  ground truth, which we  use only at test time to compare our methods. Furthermore this synthetic MRI data inherits the realistic structure of the NLLS estimations used to generate them, which themselves get their structure from the in vivo mGRE MRI data. The synthetic data was split into 23 datasets for training, 4 for validation and 8 for testing. Data from patients who were scanned twice was put in the same sets, to avoid biasing our results.

\begin{figure}[!t]
\centering \includegraphics[width=0.90\textwidth]{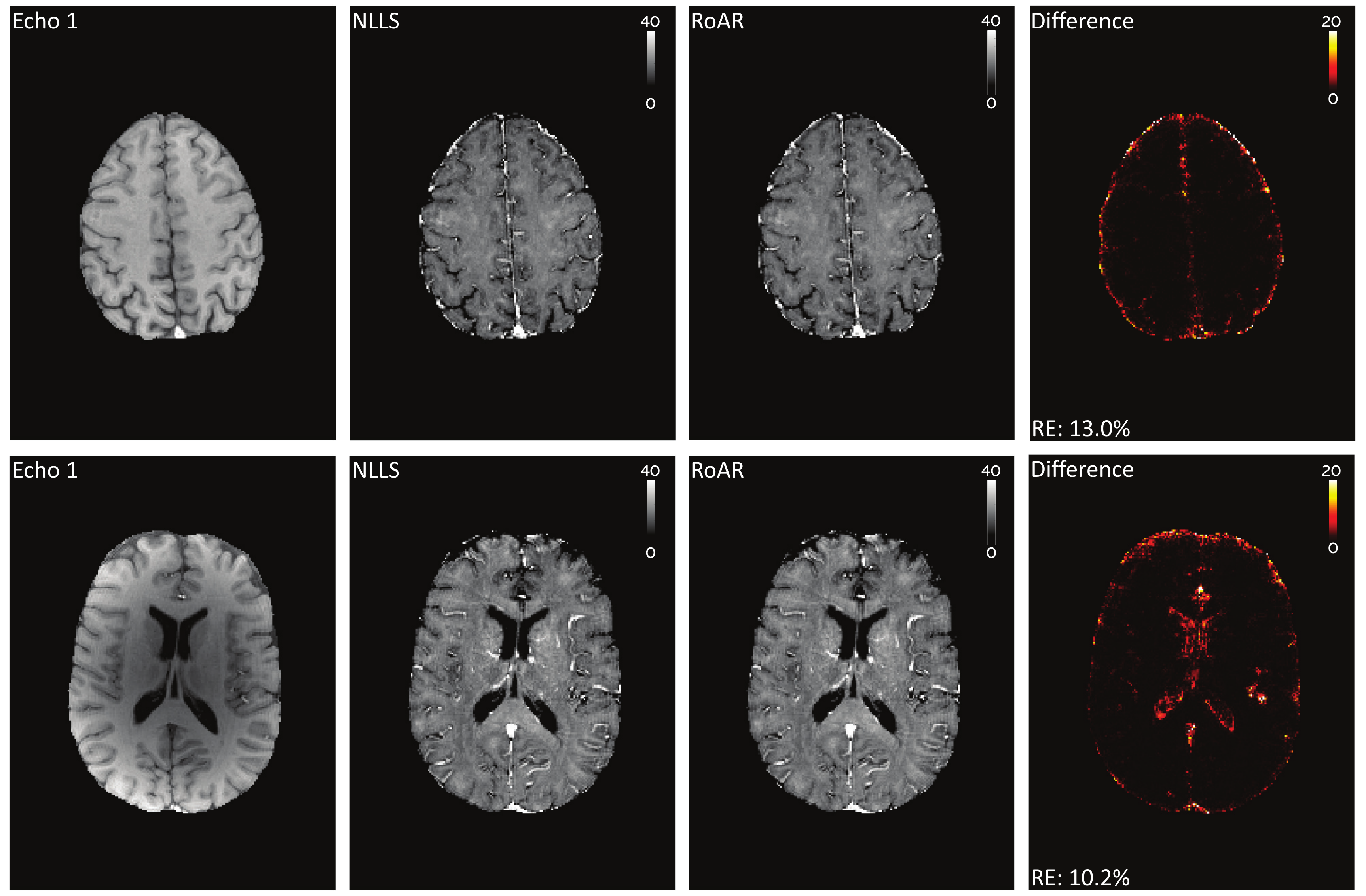}
\caption{Examples of $R_2^\ast$ maps from NLLS and RoAR methods from two in vivo slices. VSF was used to calculate NLLS results and train RoAR, but not during application of the RoAR model. The top and bottom left images show Echo 1 of the 10 input images used to produce the $R_2^\ast$ maps. The two rightmost images are maps of differences between the RoAR and NLLS results. The bottom left corner of each difference map shows the relative error between the RoAR and NLLS results.}
\label{Fig:RealData}
\end{figure}

In the synthetic training paradigm, we use copies of the simulated MRI data with added noise. This noise is added before MRI data are split into slices for training, and comes from the distribution $\mathcal{N}(0,\frac{\bar{S_0}}{\text{SNR}})$ where $\bar{S_0}$ represents the mean $S_0$ value over the entire 3D volume of a given MRI. The use of $\bar{S_0}$ ensures that noise levels are standardized across MRI data with different signal strength, and the use of $\text{SNR}$ allows for control over noise level. For each copy $\text{SNR}$ is randomly selected from the interval $[5,20]$ to make our models robust to inputs with a range of noise levels. Four copies of the synthetic data, and their noisy counterparts were used to train RoAR's U-Net for approximately $2.5$ hours. The noisy synthetic data and the ground truth $R_2^\ast$ maps used for its generation were also used to train a standard supervised network for approximately $2$ hours. Additionally, three sets were made, each consisting of copies of the 8 synthetic test MRI data, with the noise level of all MRI data in a set corresponding to either $\text{SNR} = 5$, $10$ or $15$. At test time, these sets were used to compare how robust each method is to different levels of noise. 

We use the \emph{relative error (RE)} metric as a means to quantitatively compare generated $R_2^\ast$ maps with the ground truth
\begin{equation}
    \text{RE} =  \frac{\left\|\rbf_2^\ast -  \rbfhat_2^\ast\right\|} {\left\|\rbf_2^\ast\right\|} \times 100 \%,
\end{equation}
where $\rbfhat_2^\ast$ and $\rbf_2^\ast$ represent a given methods' $R_2^\ast$ output and the ground truth, respectively. Here $\left\|\cdot\right\|$ denotes the standard euclidean norm. We note that RE is computed inside a brain mask.

\section*{Results}
\label{sec:results}

Figure~\ref{Fig:RealData} shows examples for two in vivo slices of the $R_2^\ast$ calculated by NLLS and RoAR trained on in vivo data, as described above. It can be seen that the $R_2^\ast$ maps for the two methods are both of high quality and are very similar (SNR in original mGRE data is about 50). Importantly, in the RoAR approach, the F-function was only used during training but not estimation, i.e. the final CNN model $\Ical_\thetabm$ operates in the domain of magnitude images $\sbf$ and does not need input from $F(t)$. We note that while the NLLS predictions on high SNR in vivo data are a good reference for evaluation, they should not be thought of as ground truth.

Table~\ref{tab:RelativeError} provides a summary of the average relative error of results from the methods over all \emph{brain slices} from the 8 synthetic test MRI data, given separately for each of the three noisy test sets. Here \emph{brain slices} correspond to slice 25 to 55 of a 72 slice MRI, 20 to 50 of a 60 slice MRI, and 30 to 60 when there are 88 slices. At every input noise level, RoAR and the supervised approach have much lower RE than NLLS. This quality gap grows noticeably as the input gets noisier: from around $1 \%$ at SNR=15 to around $25 \%$ at SNR=5. 

\begin{table}[H]
\centering
\caption{Average Relative Errors of $R_2^\ast$ evaluation from NLLS, RoAR and the traditional supervised approach on the synthetic test data for three different noise levels. RE were computed inside brain masks that insure removing all skull voxels where the signal model in Eq.~\eqref{Eq:Model} is not applicable}
  \label{tab:RE}

\begin{tabular}{lccc}
\hline
\textbf{Method} & \textbf{SNR=5} & \textbf{SNR=10} & \textbf{SNR=15} \\
\hline
NLLS & 46.6 \% & 23.8 \% & 15.8 \% \\
RoAR & \textbf{22.0 \%}& \textbf{17.1 \%} & \textbf{14.9 \%}\\
Supervised & 22.5 \% & 17.2 \% & \textbf{14.9 \%} \\
\hline 
\end{tabular}
\label{tab:RelativeError}
\end{table}

Figure~\ref{Fig:SNR5And10} shows examples of the RoAR results for the same two slices, from test sets with different noise levels (SNR 5 and 10). All NLLS-generated $R_2^\ast$ maps have higher noise levels when compared with the corresponding RoAR results. This is highlighted by the $R_2^\ast$ difference maps (generated $-$ ground-truth) located under each $R_2^\ast$ map, visualizing the absolute value of their deviation from the  ground truth. The NLLS difference maps are much brighter for the low SNR data and, although less extreme, are noticeably brighter then RoAR's difference maps for the higher SNR data. The RE of each estimated image is also shown in the bottom left corner. In all images shown the RE of the RoAR results are lower then NLLS's.

Figures~\ref{Fig:RealData_Supervised} and~\ref{Fig:SNR5And10_Supervised} in the Appendix show results for the traditional supervised approach on experimental and noisy synthetic data respectively. Both qualitatively and quantitatively, these images are very similar to the RoAR results in Figures~\ref{Fig:RealData} and~\ref{Fig:SNR5And10}. 

\section*{Discussion and Conclusions}
\label{sec:conclusion}


In this manuscript we proposed a self-supervised Convolutional Neural Network approach for fast and robust estimation of $R_2^\ast$ maps from a multi-Gradient-Recalled Echo MRI data. The method is based on a deep neural network that uses a biophysical model connecting MRI signal with underlying biological tissue microstructure. This approach has an advantage for model parameter estimation compared with training based on ground truth data that is not available from actual experimental data contaminated by noise and imaging artifacts. Figure~\ref{Fig:RealData} shows that RoAR has the ability to produce $R_2^\ast$ images from high SNR in vivo data of the same quality as NLLS-based voxel-by-voxel analysis. In this regard there are two important positives of using RoAR over NLLS. The first is RoAR's faster runtime:  RoAR takes 5 seconds to output $R_2^\ast$ and $S_0$ maps for a full brain (using a GeForce GTX 1080 Ti GPU) while NLLS takes 120 minutes on a modern PC (using 8 cores). The second improvement is that RoAR operates in the domain of magnitude mGRE images and is trained to recognize the contribution of macroscopic magnetic field inhomogeneities to the mGRE signal only from the magnitude data, thus providing $R_2^\ast$ maps free from macroscopic magnetic field inhomogeneity artifacts. At the same time, a standard NLLS analysis requires both magnitude and phase images to compute $F(t)$ values that are subsequently used during NLLS fitting. It is important to emphasize that pre-computed $F(t)$ values are an essential input to the biophysical model that is used to train RoAR, but are not needed by RoAR to produce $R_2^\ast$ maps after training. The input being transformed by the U-Net is just magnitude mGRE images and the output is $R_2^\ast$ maps free from macroscopic magnetic field inhomogeneity artifacts even though mGRE images are naturally affected by magnetic field inhomogeneities.

\begin{figure}[!t]
\centering \includegraphics[width=0.95\textwidth]{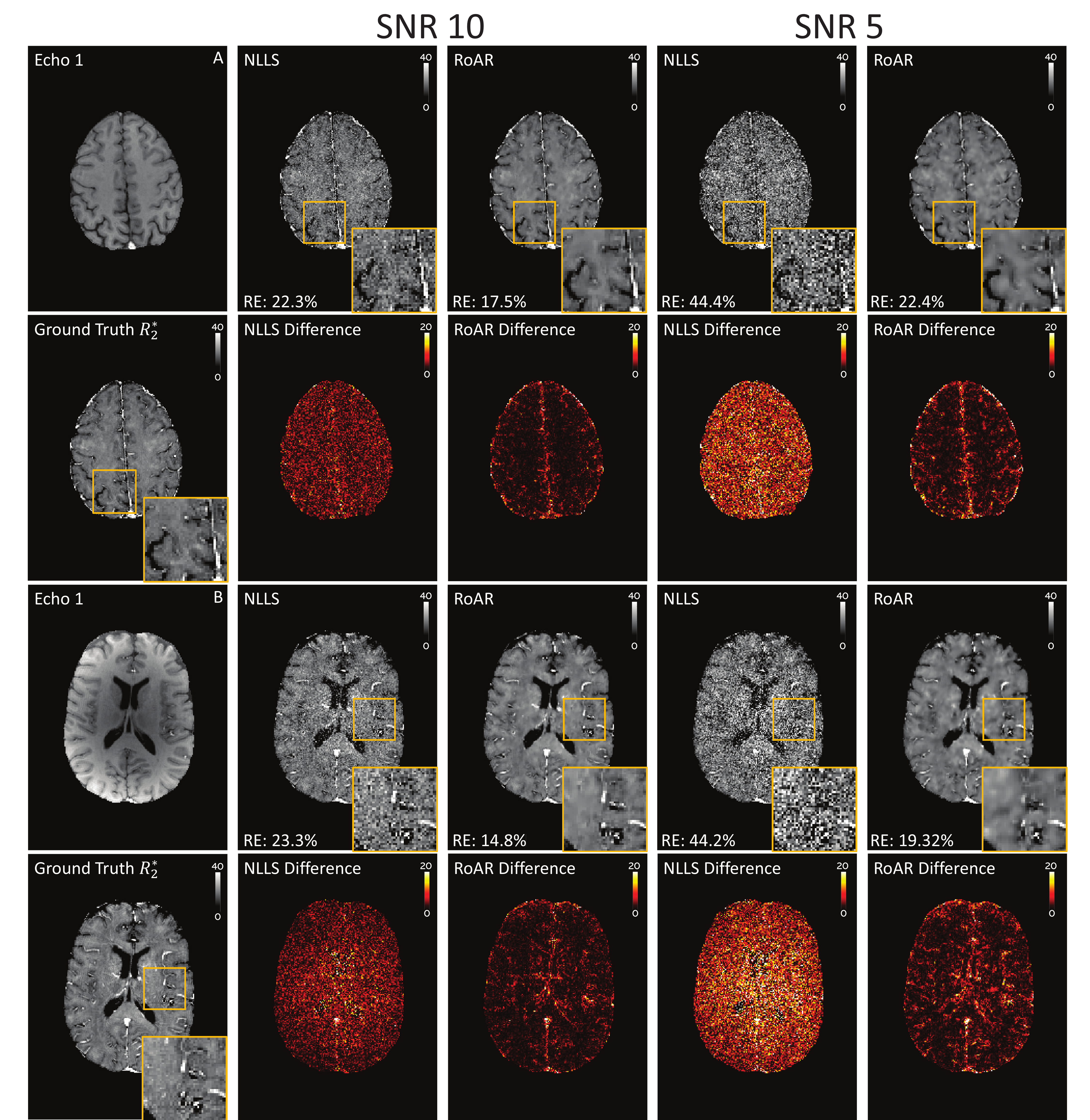}
\caption{Denoising feature of the RoAR model. Images represent two examples of the results obtained from the synthetic datasets with different noise levels (SNR 5 and 10). Images of Echo 1 $(\sbf_1)$ of the method input (before the addition of noise) and the ground truth $R_2^\ast$ maps are shown in the first column. Columns 2 and 3 are $R_2^\ast$ maps from input with SNR 10, while columns 4 and 5 correspond to those from input with SNR 5. The RE of each $R_2^\ast$ map is shown in its bottom left corner. Representative regions with $2\times$ zoom are shown on the bottom right of each $R_2^\ast$ map. Underneath each $R_2^\ast$ map is its difference with the ground truth.}
\label{Fig:SNR5And10}
\end{figure}

Another important advantage of RoAR is its denoising feature. The synthetic MRI experiments show that RoAR is significantly more robust than NLLS with respect to noise in the data: as illustrated by Table~\ref{tab:RelativeError} and Figure~\ref{Fig:SNR5And10}, the Relative Error in $R_2^\ast$ evaluation is significantly smaller in RoAR-estimated $R_2^\ast$ data than in the NLLS results. This is  due to two limitations in NLLS. While the RoAR estimations are based on the data from the entire image, NLLS is a voxel-based approach. Because of this NLLS cannot take advantage of the naturally occurring spatial patterns in the images that RoAR benefits from. The second disadvantage of NLLS is that RoAR was trained to take noisy signal and output $R_2^\ast$ corresponding to cleaner signal, while NLLS is simply run directly on the noisy data.  This further limits NLLS output quality as it never gets to ''see'' examples of what high SNR data looks like.



It is clear from the synthetic results that the relative improvement of $R_2^\ast$ estimates by RoAR over NLLS grows as the input gets noisier. In Table 1 the difference between RoAR and NLLS's RE changes from $0.9 \%$ to $6.7 \%$ when going from SNR=15 to SNR=10 and shoots up to $24.6 \%$ when going from SNR=10 to SNR=5. The results in Figure~\ref{Fig:SNR5And10} qualitatively show this relative quality increase. While also strong with lower noise input, RoAR becomes an increasingly attractive option as noise increases.

We would also like to note that training a universal RoAR that could be applied to any sequence parameters is probably not a realistic task. The goal of our paper is only to demonstrate the feasibility of training based on a biophysical model. Such training can be done for any sequence parameters and the resultant RoAR can be implemented on the scanner with a given sequence with the goal of generating R2* maps free from field inhomogeneity artifacts in a matter of seconds.

\textbf{In Conclusion,} we introduced RoAR as a fast, self-supervised, method that can utilize  magnitude only mGRE data to produce high quality $R_2^\ast$ maps free from artifacts resulting from macroscopic magnetic field inhomogenieties. As a self-supervised network, RoAR training does not rely on the NLLS-estimated $R_2^\ast$ maps that are needed as a ground truth for traditional supervised training. We also demonstrated  RoAR's ability to estimate $R_2^\ast$ maps with high accuracy even from noisy mGRE signals with magnetic field inhomogenieties. This can be achieved because, as a CNN based approach, RoAR utilizes information on the signal not just from individual voxels (as in a voxel-based NLLS approach) but also spatial patterns of the signals in the images. Future work will consider the modification of RoAR to take in 3D images with echos, allowing for the leveraging of more complex 3D spatial statistical patterns in estimations. The resilience of RoAR to the noise in the data, and high estimation speed provide significant advantages of using RoAR in clinical settings.

\section*{Appendix}
\label{sec:appendix}
\begin{figure}[H]
\centering \includegraphics[width=0.90\textwidth]{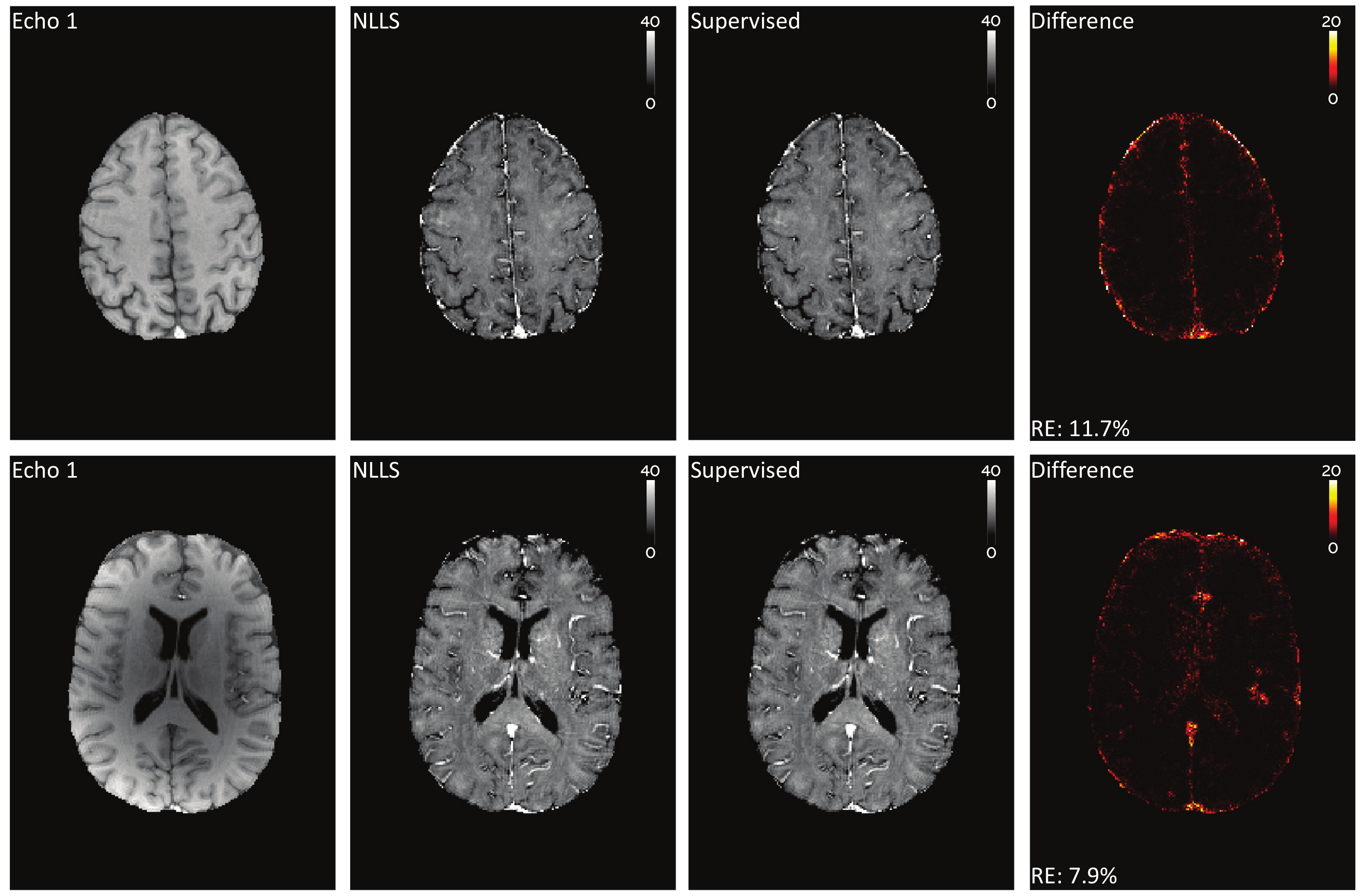}
\caption{Examples of $R_2^\ast$ maps obtained from NLLS and the traditional supervised approach from two in vivo slices. NLLS-generated $R_2^\ast$ maps were used as ground truth for supervised training. The top and bottom left images show Echo 1 of the 10 input images used to produce these $R_2^\ast$ maps. The two rightmost images are maps of differences between the NLLS and supervised results. The bottom left corner of each difference map shows the relative error between the supervised and NLLS results.}
\label{Fig:RealData_Supervised}
\end{figure}

\begin{figure}[H]
\centering \includegraphics[width=0.95\textwidth]{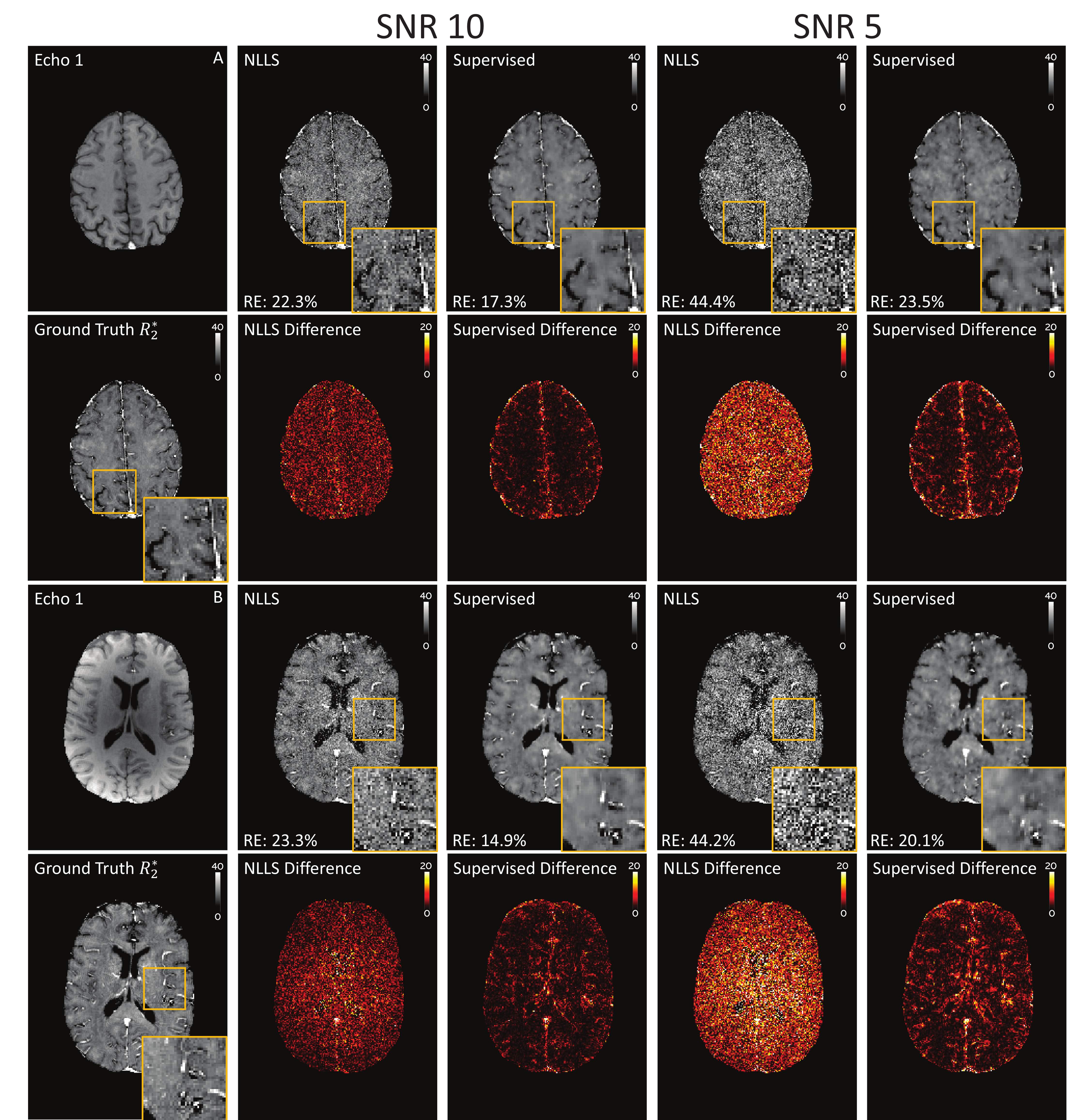}
\caption{Two examples of the results obtained from the synthetic datasets with different noise levels (SNR 5 and 10). Images of Echo 1 $(\sbf_1)$ of the method input (before the addition of noise) and the ground truth  $R_2^\ast$ are shown in the first column. Columns 2 and 3 correspond to $R_2^\ast$ maps from input with SNR 10, while columns 4 and 5 correspond to those from input with SNR 5. The RE of each $R_2^\ast$ map is shown in its bottom left corner. Representative regions with $2\times$ zoom are shown on the bottom right of each $R_2^\ast$ map. Underneath each $R_2^\ast$ map is its difference with the ground truth.}
\label{Fig:SNR5And10_Supervised}
\end{figure}


\textbf{Acknowledgement:} The authors are grateful to Vadim Omeltchenko, Sr.~AWS Solutions Architect, for helpful discussion.

\clearpage
\listoffigures

\end{document}